\documentclass[aps,twocolumn,showpacs,superscriptaddress]{revtex4}
\usepackage{graphicx}
\usepackage{times}

\begin{document}

\title{Competition and cooperation among different punishing strategies in the spatial public goods game}

\author{Xiaojie Chen}
\email{xiaojiechen@uestc.edu.cn}
\affiliation{School of Mathematical Sciences, University of Electronic Science and Technology of China, Chengdu 611731, China}

\author{Attila Szolnoki}
\email{szolnoki@mfa.kfki.hu}
\affiliation{Institute of Technical Physics and Materials Science, Centre for Energy Research, Hungarian Academy of Sciences, P.O. Box 49, H-1525 Budapest, Hungary}

\author{Matja{\v z} Perc}
\email{matjaz.perc@uni-mb.si}
\affiliation{Faculty of Natural Sciences and Mathematics, University of Maribor, Koro{\v s}ka cesta 160, SI-2000 Maribor, Slovenia}
\affiliation{Department of Physics, Faculty of Sciences, King Abdulaziz University, Jeddah, Saudi Arabia}
\affiliation{CAMTP -- Center for Applied Mathematics and Theoretical Physics, University of Maribor, Krekova 2, SI-2000 Maribor, Slovenia}

\begin{abstract}
Inspired by the fact that people have diverse propensities to punish wrongdoers, we study a spatial public goods game with defectors and different types of punishing cooperators. During the game, cooperators punish defectors with class-specific probabilities and subsequently share the associated costs of sanctioning. We show that in the presence of different punishing cooperators the highest level of public cooperation is always attainable through a selection mechanism. Interestingly, the selection not necessarily favors the evolution of punishers who would be able to prevail on their own against the defectors, nor does it always hinder the evolution of punishers who would be unable to prevail on their own. Instead, the evolutionary success of punishing strategies depends sensitively on their invasion velocities, which in turn reveals fascinating examples of both competition and cooperation among them. Furthermore, we show that under favorable conditions, when punishment is not strictly necessary for the maintenance of public cooperation, the less aggressive, mild form of sanctioning is the sole victor of selection process. Our work reveals that natural strategy selection can not only promote, but sometimes also hinder competition among prosocial strategies.
\end{abstract}

\maketitle

\section{Introduction}
Cooperation is vital for the maintenance of public goods in human societies \cite{nowak_s06, nowak_11}. But according to Darwin's theory of evolution, competition rather than cooperation ought to drive our actions. The reconciliation of this theory with the fact that cooperation is widespread in human societies, as well as with the fact that it is much more common in nature as one might expect, is one of the most persistent challenges in evolutionary biology and social sciences \cite{pennisi_s05, colman_n06, nowak_jtb12, rand_tcs13}. Past decades have seen the paradigm of punishment rise as one of the more successful strategies by means of which cooperation might be promoted \cite{fehr_n02, brandt_prsb03, helbing_njp10, boyd_s10, baldassarri_pnas11, sasaki_pnas12, perc_jrsi13, vasconcelos_ncc13, cui_pb_jtb14}. Indeed, punishment is also the principle tool of institutions in human societies for maintaining cooperation and otherwise orderly behavior \cite{henrich_s06, gurerk_s06, sigmund_n10, szolnoki_pre11}. However, punishment is costly, and as such it reduces the payoffs of both the defectors as well as of those that exercise the punishment, hence yielding an overall lower income and acting as a drain on social welfare. Thus, understanding the emergence of costly punishment is crucial for the evolution of cooperation \cite{fowler_n05, hauert_s07, milinski_n08, helbing_ploscb10, szolnoki_pre11b, hilbe_pnas14, hintze_pb15, chen_xj_njp14}.

While recent research confirms that punishment is often motivated by negative personal emotions such as anger or disgust \cite{fehr_n02, cubitt_ee11}, Raihani and McAuliffe have shown also that the decision to punish is often motivated with the aversion of inequity in mind, rather than by the desire for reciprocity \cite{raihani_bl12}. Although prosocial punishment is widespread in nature \cite{sigmund_tee07, shimao_pone13}, it is unlikely that cooperators are willing to commit permanently to punishing wrongdoers. For that, the action is simply to costly, and hence some form of abstinence is likely, also to avoid unwanted retaliation. Several research groups have recently investigated these and related up and down sides of punishment \cite{janssen_jtb08, rand_jtb09, perc_njp12, wolff_jtb12, szolnoki_jtb13, chen_xj_njp14}. For example, it was shown that cooperators punish defectors selectively depending on their current personal emotions, even if the number of defectors is large \cite{raihani_bl12}. More often than not, however, whether or not to punish depends on the whiff of the moment and is thus a fairly random event. Motivated by these observations, we have recently shown that sharing the effort of punishment in a probabilistic manner can significantly lower the vulnerability of costly punishment and in fact help stabilize costly altruistic strategies \cite{chen_xj_njp14}.

Here we drop the assumption that cooperators who do punish defectors do so uniformly at random. Instead, we account for the diversity in punishment, taking into account the fact that some individuals are more likely to punish, while others punish only rarely. More specifically, we introduce different threshold levels for punishment, which ultimately introduces different classes of cooperators that punish defectors. The assumption of diverse players is not just a realistic hypothesis, but in general it is firmly established that it also has a decisive impact on the evolution of public cooperation \cite{perc_pre08, santos_n08, perc_njp11, santos_jtb12}. Motivated by this fact, we therefore study a spatial public goods game with defectors and different types of punishing cooperators. While previously we have demonstrated the importance of randomly shared punishment \cite{chen_xj_njp14}, we here approach a more realistic scenario by assuming that each type of cooperators will punish with a different probability. Our goal is to determine whether a specific class of punishing cooperators will be favored by natural selection, or whether despite the competition among them synergistic effects will emerge. As we will show, the evolution is governed by a counterintuitive selection mechanism, depending further on the synergistic effects of cooperative behavior. However, before presenting the main results in detail, we proceed by a more accurate description of the studied spatial public goods game with different punishing strategies.

\section{Spatial public goods game with diverse punishment}
We consider a population of individuals who play the public goods game on a square lattice of size $L \times L$ with periodic boundary conditions. We assume that the game is contested between $T$ classes of cooperators ($C_0$, $C_1$, $\ldots$, $C_{T-1}$) and defectors ($D$). Independently of the class a cooperator belongs to, it contributes an amount $c$ to the common pool, while defectors contribute nothing. After the sum of all contributions in the group is multiplied by the enhancement factor $r>1$, the resulting amount is shared equally among all group members.

Moreover, cooperators with strategy $C_i$ ($0\leq i\leq T-1$) choose to punish defectors with a probability $i/(T-1)$ if the latter are present. As a result, each defector in the group is punished with a fine $\alpha$, while all the cooperators who participated in the punishment equally shared the associated costs. In particular, each punishing cooperator bears the cost $(n-n_C)\alpha/n_P$, where $n_C$ and $n_P$ are the number of cooperators and punishers in the group, respectively. We emphasize that a cooperator who decides to punish bears the same cost independently of the class it belongs to. Thus, here the strategy $s=C_i$ only determines how frequently a cooperator is willing to punish defectors. Nevertheless, it is worth pointing out that $C_0$ never punish and thus correspond to traditional second-order free-riders because they enjoy the benefits of punishment without contributing to it \cite{panchanathan_n04}. On the other extreme, cooperators belonging to the $C_{T-1}$ class punish always when defectors are present in the group. Since each player on site $x$ with von Neumann neighborhood is a member of five overlapping groups of size $N=5$, in each generation it participates in five public goods games and obtains its total payoff $P_x=\sum_j P_x^{j}$, where $P_x^{j}$ is the payoff gained from group $G_j$.

Subsequently, a player $x$, having strategy $s_x$, adopts the strategy $s_y$ of a randomly chosen neighbor $y$ with the probability
\begin{equation}
f(s_x \leftarrow s_y)=\frac{1}{1+\exp[(P_x-P_y)/\kappa]},
\end{equation}
where $\kappa$ denotes the amplitude of noise \cite{szabo_pr07}. Without loosing generality and to ensure continuity of this line of research \cite{szolnoki_pre09c} we set $\kappa=0.5$, meaning that it is very likely that the better performing players will pass their strategy to their neighbors, yet it is also possible that players will occasionally learn from a less successful neighbor. To conclude the description of this public good game, we would like to emphasize that different $C_i$ classes represent different strategies, as our goal is to explore how the willingness to punish evolves at specific parameter values.

The model is studied by means of Monte Carlo simulations. Initially, defectors randomly occupy half of the square lattice, and each type of cooperators randomly $1/T$ of the rest of
the lattice. During one full Monte Carlo step (MCS), all individuals in the population receive a chance once on average to adopt another strategy. Depending on the proximity to phase transition points and the typical size of emerging spatial patterns, the linear system size was varied from $L=120$ to $600$ and the relaxation time was varied from $10^4$ to $10^6$ MCS to ensure proper statistical accuracy. The reported fractions of competing strategies were determined in the stationary state when their average values became time-independent. Alternatively, we have averaged the outcomes over $20-100$ independent runs when the system terminated into a uniform absorbing state.

\section{Results}

\begin{figure}
\centerline{\includegraphics[width=7cm]{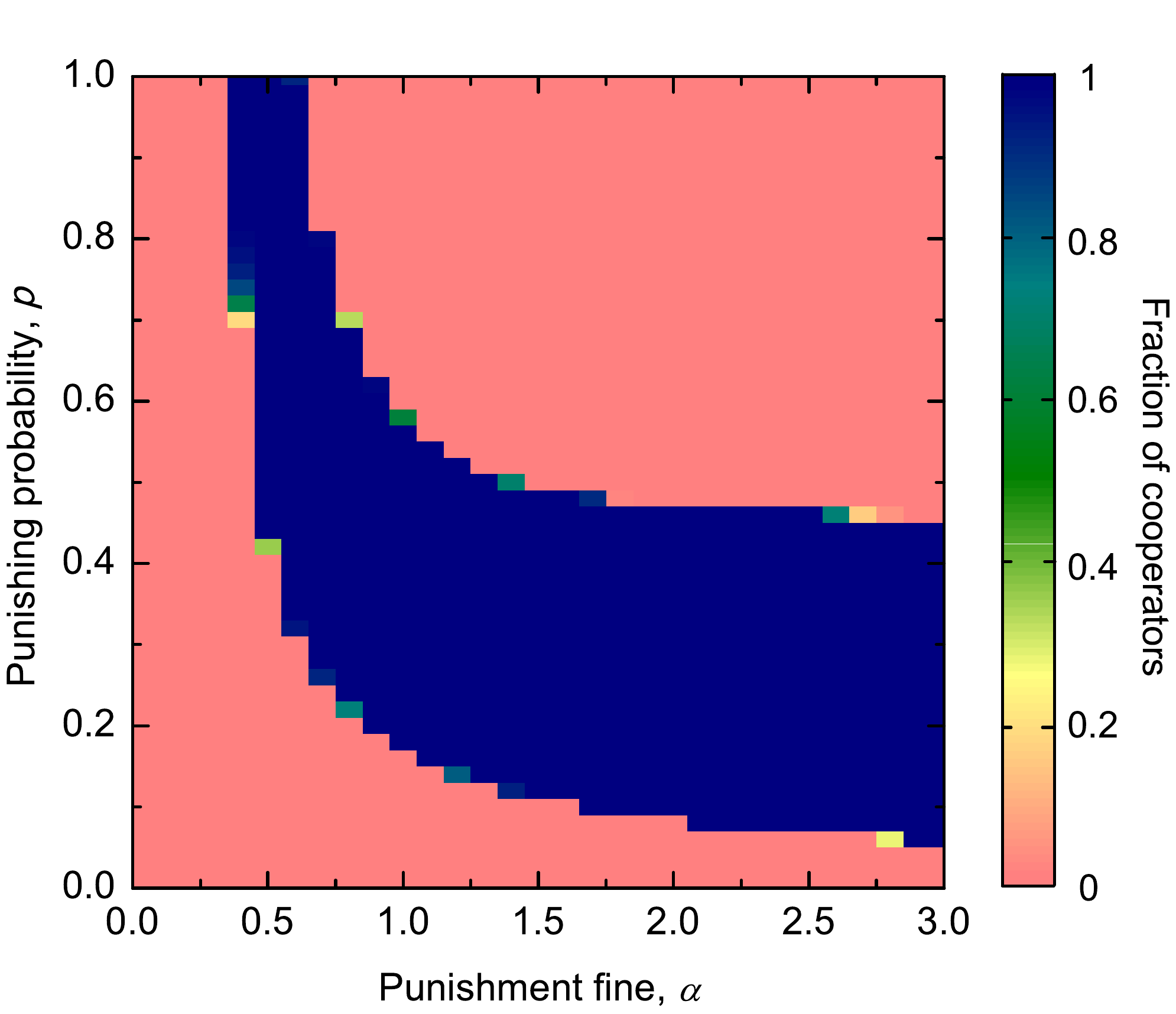}}
\caption{(Color online) Fraction of cooperators as a function of the punishment fine $\alpha$ and the probability to punish $p$, as obtained for a low multiplication factor $r=3.5$ in the original model proposed in \cite{chen_xj_njp14}, where a uniform probability to punish was assumed for all cooperators. Note that both $\alpha$ and $p$ have a non-monotonous impact on the fraction of cooperators.}
\label{fig1}
\end{figure}

For the sake of comparison, we first present the fraction of cooperators in dependence on the punishment fine $\alpha$ and the probability to punish $p$ at a low $r$ value, as obtained in the original probabilistic punishment model, where cooperators punish uniformly at random \cite{chen_xj_njp14}. Figure~\ref{fig1} illustrates that the fraction of cooperators first increases, reaches its maximum, but then again decreases, as the values of $\alpha$ and $p$ increase along the diagonal on the $p-\alpha$ plane. Increasing one of these parameters, while the other is kept constant, returns to the same observation. Both $\alpha$ and $p$ thus have a non-monotonous impact on the fraction of cooperators, which is closely related with the fact that $\alpha$ characterizes not only the level of punishment but also its cost. Accordingly, too high values of $\alpha$ involve too high costs stemming from the act of punishing. It is worth pointing out that $r=3.5$, which is used in Fig.~\ref{fig1}, is a relatively low value of the multiplication factor at which the non-monotonous dependence can still be observed. In comparison with the results obtained for larger values of $r$ as used in Ref.~\cite{chen_xj_njp14}, however, the current plot features a significantly narrower $p$ region where full cooperation is possible when $\alpha$ is sufficiently large. Similarly, there is a limited region of intermediate $\alpha$ values where cooperators that punish severely can beat defectors. Based on these observations, in the present model we thus explore if there is an evolutionary selection among different punishing strategies as they compete against the defectors simultaneously, or if there is indeed cooperation in the common goal to deter defectors.

\begin{figure}
\centerline{\includegraphics[width=8.5cm]{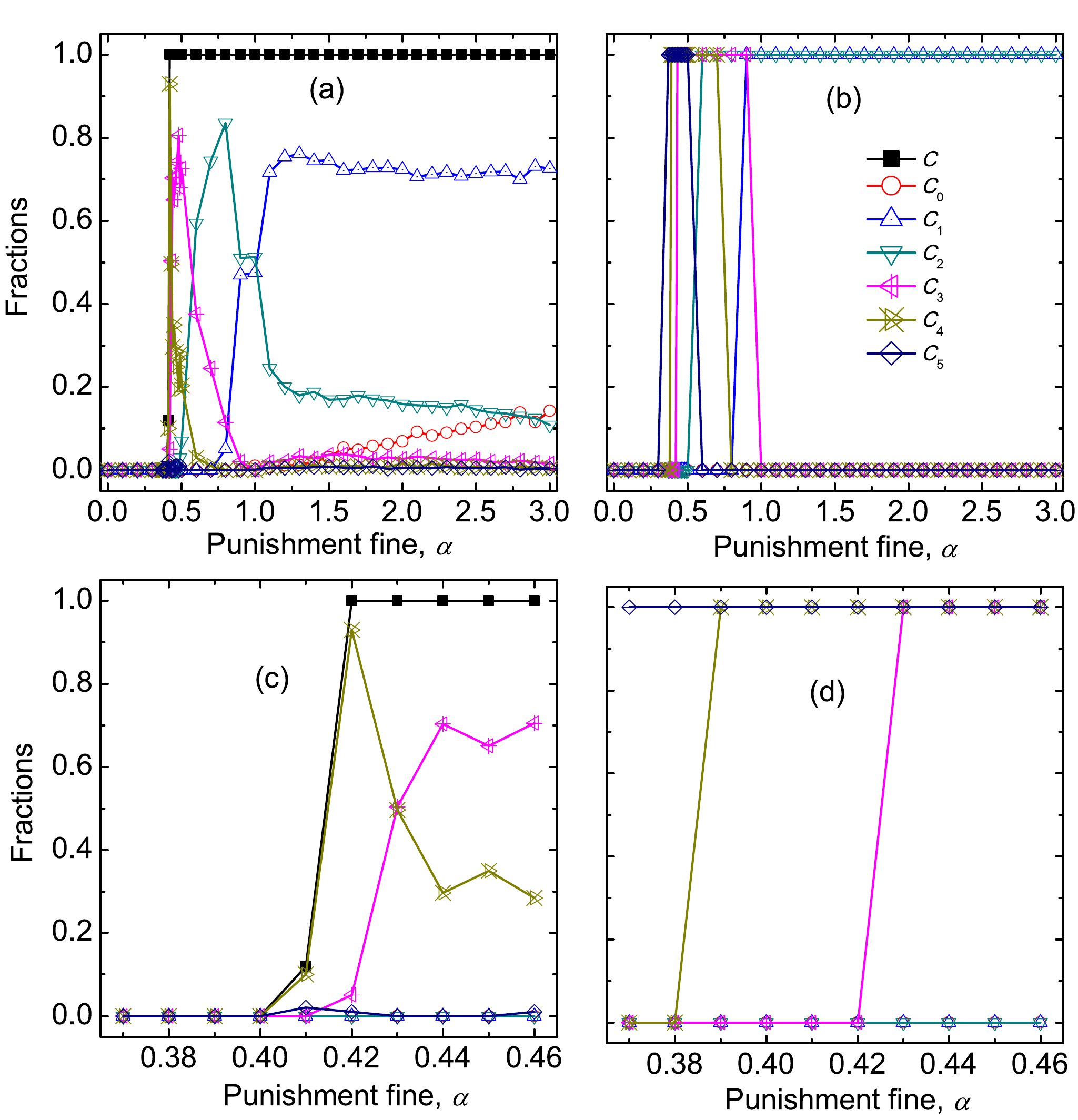}}
\caption{(Color online) Panel~(a) shows the fraction of different cooperator classes in the final state in dependence of $\alpha$ when they start fighting with defectors simultaneously. Panel~(c) shows an enlarged part of panel~(a) at low $\alpha$ values, when cooperation becomes dominant over defection. To present the overall level of cooperation in the population, the cumulative fraction of $C_i$ strategies is also shown (denoted by $C$). For comparison, in panel~(b) we have also plotted the resulting fraction of cooperator classes when they fight against defectors individually. As in panel (c), panel~(d) shows an enlarged part of panel~(b) at a specific interval of $\alpha$. The multiplication factor in all panels is $r=3.5$.}
\label{fig2}
\end{figure}

For an intuitive overview, we set $T=6$ and investigate how the six types of punishing strategies compete and potentially cooperate with each other in the presence of defectors. The general conclusion, however, is robust and remains valid if we use other values of $T$. Using the same $r=3.5$ as in Fig.~\ref{fig1}, the panels of Fig.~\ref{fig2} summarize our main findings. The first panel shows the fractions of strategies in the final state in dependence of the punishment fine
$\alpha$ when different punishing strategies fight against defectors simultaneously. For clarity, we have also plotted the accumulated fraction of punishing strategies. In contrast to the uniform punishing model, we can see that the total fraction of cooperators should increase monotonously with increasing $\alpha$. As Fig.~\ref{fig2}(a) illustrates, cooperators can survive when $\alpha>0.40$, and become dominant over $\alpha \geq 0.42$ (see also the enlarged part in Fig.~\ref{fig2}(c)). We should stress, however, that not all types of cooperators can survive at equilibrium, even if cooperators take over the whole population. It turned out that there are some ``weak'' classes of cooperators who go extinct before defectors die out, while other classes of cooperators survive.

For a more in-depth explanation, the vitality of punishing classes can be estimated if we let them fight against defectors individually. The outcomes of this scenario are summarized in Fig.~\ref{fig2}(b). Results presented in this panel suggest that there are punishing classes who can dominate for all high $\alpha$ values, while others become vulnerable as we increase $\alpha$. More interestingly, however, there are mildly punishing strategies who can survive only due to the support of the more successful punishing strategies. For example, for $\alpha=0.42$ classes $C_5$ and $C_4$ can outperform defectors, while $C_3$ disappear when they fight against defectors individually [Fig.~\ref{fig2}(b) and (c)]. But when all punishing strategies are  on the stage then $C_3$ players can survive as well. This effect is more spectacular for the second-order free riding $C_0$ class, who would die out immediately at such a low synergy factor $r$ if they face defectors alone. But now, especially at high $\alpha$ values, their ratio becomes considerable. This indicates that some less viable classes of cooperators can survive because of the support of more viable punishing strategies via an evolutionary selection mechanism which has a biased impact on the evolution of otherwise competing strategies.

\begin{figure}
\centerline{\includegraphics[width=8.5cm]{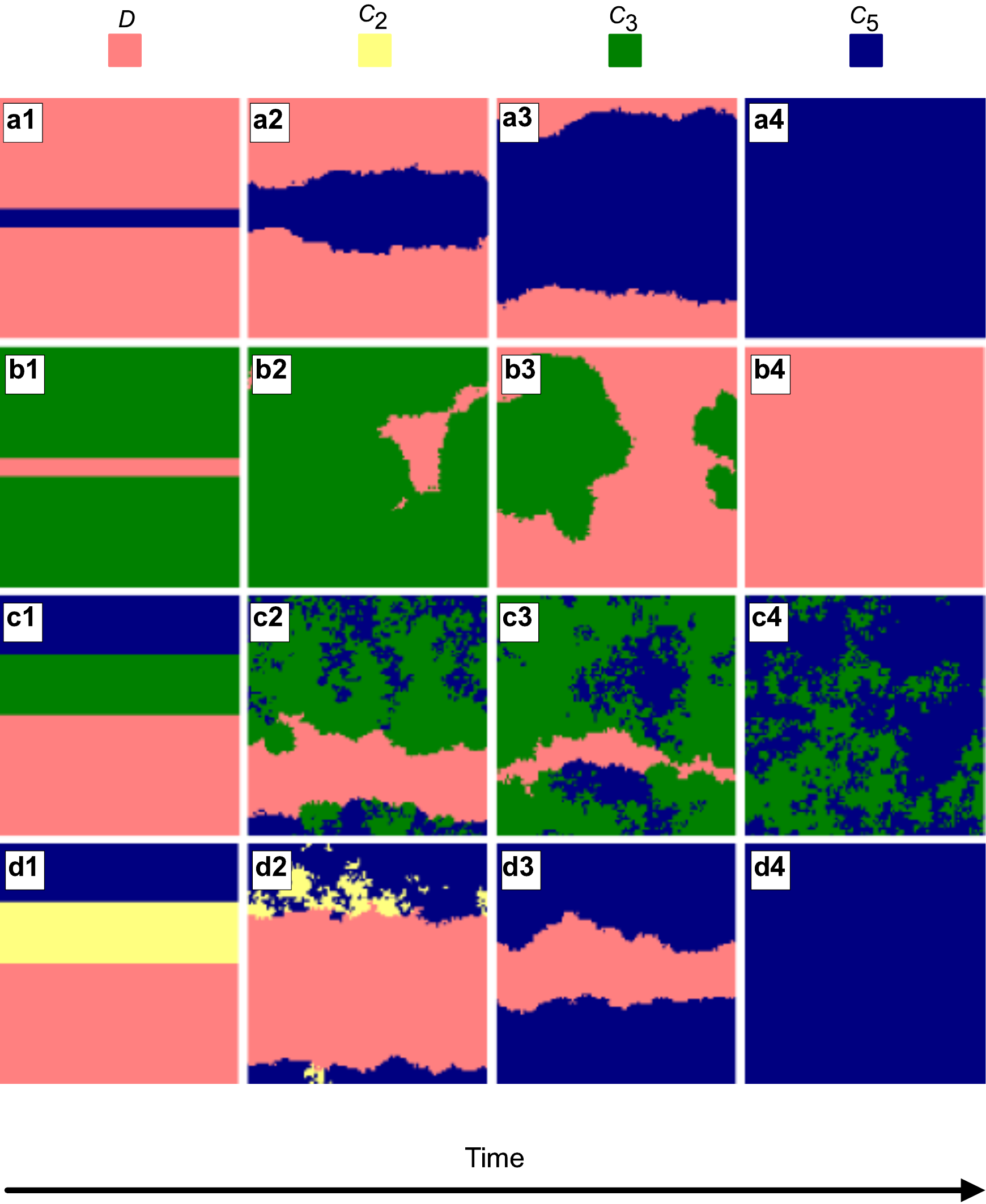}}
\caption{(Color online) Evolution of typical spatial patterns, as obtained for four different prepared initial conditions when using $\alpha=0.42$ and $r=3.5$. The first row shows the case when just a few $C_5$ cooperators are initially present among defectors. It can be observed that even under such unfavorable initial conditions the $C_5$ strategy can successfully outperform defectors. The second row feature a similar experiment with the $C_3$ strategy, which fails to survive among defectors even though the latter are initially in minority. The third row illustrates cooperation among strategies $C_3$ and $C_5$, which together dominate the whole population even though $C_3$ alone would fail under the same conditions (see second row). We note that a neutral drift starts when defectors die out, as explained in the main text. The fourth row demonstrates, however, that the cooperation among different punishing strategies illustrated in the third row is rather fragile. If initially the strategy $C_3$ is replaced by strategy $C_2$, then the later simply die out and subsequently the whole evolution becomes identical to the one shown in the first row, where strategy $C_5$ alone outperforms all defectors. For clarity, here the employed system size is small with just $L \times L = 100 \times 100$ players.}
\label{fig3}
\end{figure}

To demonstrate the underlying mechanism behind the above observations, we present a series of snapshots of strategy evolutions starting from different prepared initial states. The comparative analysis is plotted in Fig.~\ref{fig3}, where all runs were obtained for $\alpha=0.42$ and $r=3.5$. In the first row, we demonstrate how the class of $C_5$ punishing strategy can prevail over defectors. Initially, only a tiny portion of $C_5$ cooperators is launched in the sea of defectors [the fraction of $C_5$ is $8\%$, see panel (a)]. Still, $C_5$ cooperators can expand gradually and invade the whole available territory [shown in panels from (a2) to (a4)]. The second row, which was taken at the same parameter values, demonstrates clearly the vulnerability of the $C_3$ class against defectors. Despite of the fact that they occupy the majority of the available room at the beginning, shown in panel~(b1), still, they will be gradually crowded out by defector players. The final state, shown in panel~(b4), highlights that such a rare punishment activity represented by $C_3$ class is ineffective against defectors at the applied synergy factor $r$. The third row, where all previously mentioned strategies are present at the beginning, illustrates a completely different scenario. Here we start from a balanced initial state where half of the lattice sites is occupied by $C_3$ and $C_5$ strategies, while the other half is filled by defectors. As panels~(c1) to (c4) illustrate, defectors will gradually go extinct while ``weak'' $C_3$ cooperators survive and occupy almost half of the available territory in the final state. We note that there is a neutral drift between punishing strategies in the absence of defectors, which will result in a homogeneous state where the probability to arrive to one of the possible final destinations is proportional to the initial portion of a specific class at the time defectors die out \cite{cox_ap86}. This evolutionary outcome indicates that although $C_3$ players are, as an isolated strategy, weak against defector players, they can nevertheless survive because of the assistance of the strong $C_5$ strategy even if the initial fraction of the later is modest. In the fourth row, however, when we arrange a similar setup but replaced weak $C_3$ players with also weak $C_2$ players, the final state will always be the full $C_5$ state. Here, the presence of strong $C_5$ players does not yield a relevant support to $C_2$ players who therefore die out, and subsequently the system returns to the scenario illustrated in panels~(a1) to (a4).

\begin{figure}
\centerline{\includegraphics[width=8.5cm]{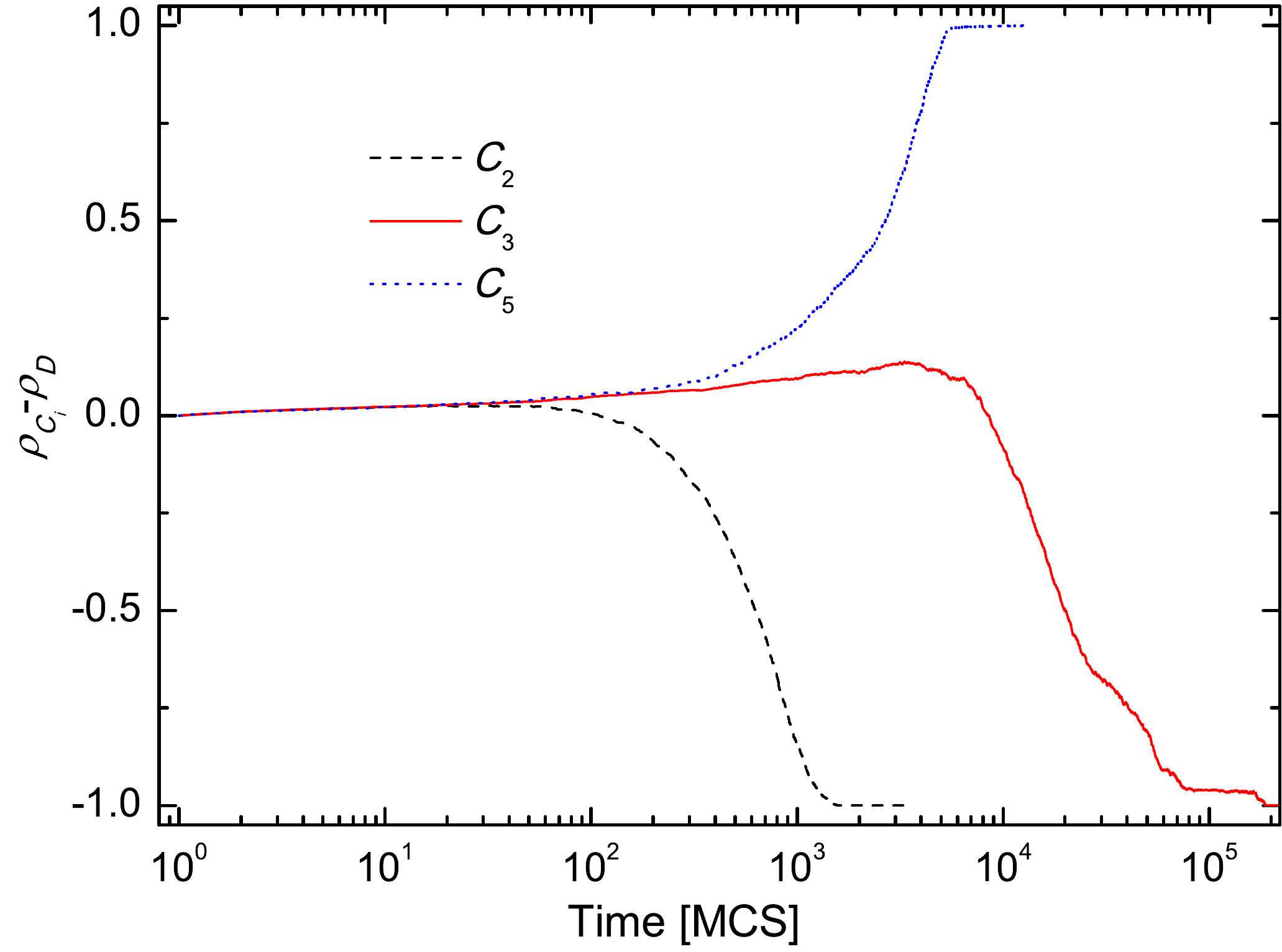}}
\caption{(Color online) Individual competition of three different punishing strategies, namely $C_2$, $C_3$ and $C_5$, against defectors in dependence on time. Note that initially only one cooperative strategy and defectors are present, using the same initial conditions as illustrated in Fig.~\ref{fig3}. Positive value of $\rho_{C_i} - \rho_D$ indicates the invasion of cooperator strategy while its negative value suggests invasion to the reversed direction. Note that while both $C_2$ and $C_3$ strategies ultimately loose their battle, the latter is able to prevail significantly longer. This enables an effective help of strategy $C_5$ when they compete against defectors together, as illustrated in panels~(c1) to (c4) in Fig.~\ref{fig3}.}
\label{fig4}
\end{figure}

The key point, which explains the significantly different trajectories for mildly punishing strategies is based on the difference of invasion velocities between the competing strategies. To demonstrate the importance of invasion velocities, we monitor how the fraction of strategies evolves in time when we launch the system from a two-strategy state where both strategies form compact domains. Following the previously applied approach illustrated in Fig.~\ref{fig3}, we compare the strategy invasions between $C_2-D$, $C_3-D$, and between $C_5-D$ strategies. The comparison of these different cases is plotted in Fig.~\ref{fig4}. As expected, both $C_2$ and $C_3$ loose the lonely fight against defectors, while $C_5$ will eventually crowd out defectors. Note that there is only a very slight increase during the early stages of the evolutionary process that can be observed for all cases, independently of the final outcome. This is because straight initial interfaces can provide a strong temporary phalanx for every punishing strategy. Nevertheless, when this interface becomes irregular due to invasions the individual weakness of $C_2$ and $C_3$ strategies reveals itself. Still, there is a significant difference between their trajectories. Namely, strategy $C_3$ is able to resist for a comparatively long time, which gives strategy $C_5$ enough time to crowd out defectors. On the other hand, strategy $C_2$ is a too easy prey for defectors, which is why they die out faster than the strategy $C_5$ is able to eliminate all defectors. Ultimately thus, strategy $C_3$ can benefit from cooperation with strategy $C_5$, while strategy $C_2$ is unable to do the same.

\begin{figure*}
\centerline{\includegraphics[width=13cm]{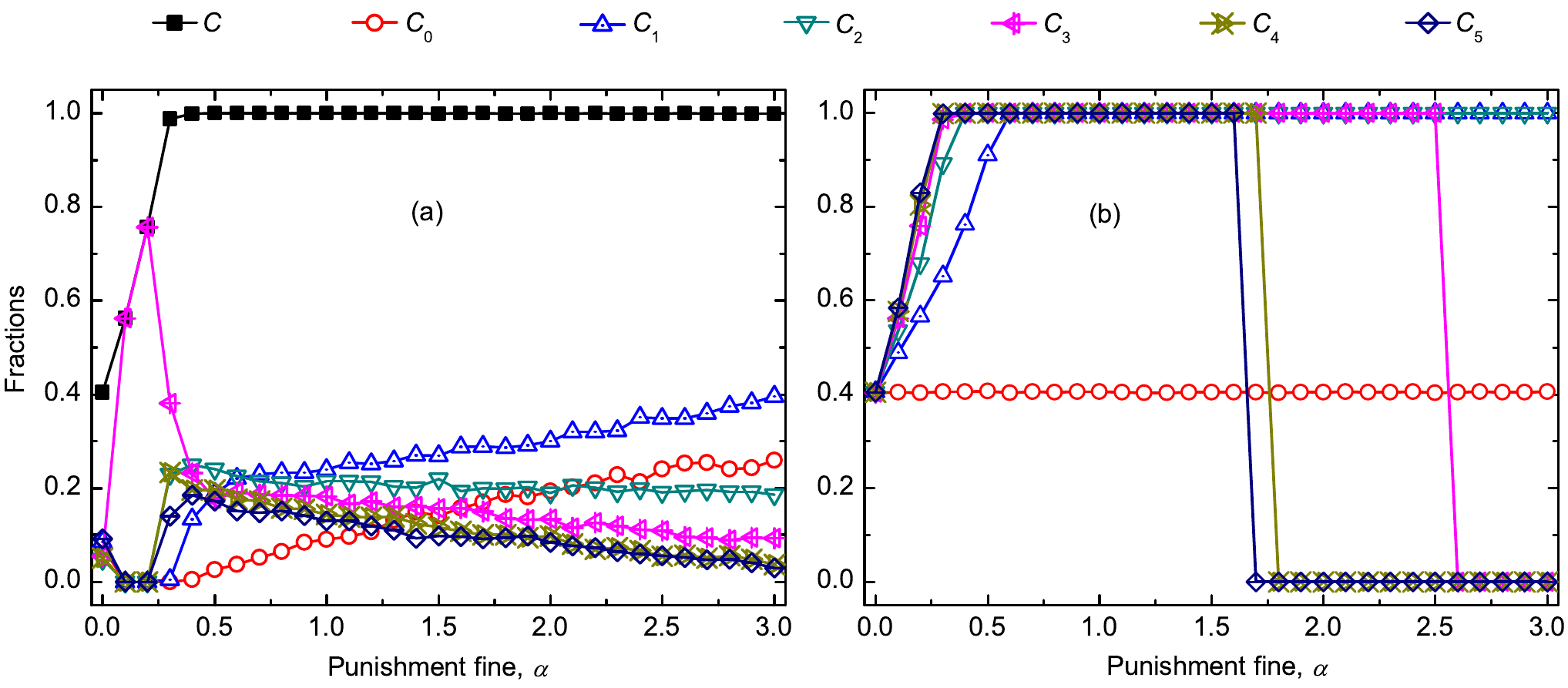}}
\caption{(Color online) Stationary fractions of different cooperator classes in dependence on $\alpha$ when they compete against defectors simultaneously [panel~(a)] and individually [panel~(b)]. The cumulative fraction of all punishing strategies (denoted as $C$ in the legend) is also plotted. The multiplication factor in both panels is $r=4.0$, which enables pure cooperators ($C_0$) to coexist with defectors even in the absence of punishment.}
\label{r4}
\end{figure*}

In the remainder of this work, we focus on the parameter region where cooperators are able to coexist with defectors without applying punishment. Namely, if the synergy factor exceeds $r>3.74$, then pure cooperators (cooperators that do not punish) can survive permanently alongside defectors due to network reciprocity \cite{szolnoki_pre09c}. Evidently, the presence of punishers can of course still elevate the overall cooperation level and defectors can be effectively crowded out from the population \cite{helbing_ploscb10}. Here the main question is thus how the different punishing strategies will share the available space.

The results are summarized in the left panel of Fig.~\ref{r4}, as obtained for the representative value of $r=4.0$. It can be observed that, when all the different types of punishing strategies fight against defectors simultaneously, then cooperators can dominate the whole population above a threshold value $\alpha>0.25$. However, to evaluate these final outcomes adequately, we need to know the individual relations between each particular cooperative strategy and defectors on a strategy-versus-strategy basis. Therefore, as for the previously presented low $r$ case in Fig.~\ref{fig2}, in the right panel of Fig.~\ref{r4} we also show the stationary fractions of different cooperators classes when they compete against defectors individually. Results presented in panel (b) highlight that too large $\alpha$ values could be detrimental for the $C_3$, $C_4$ and the $C_5$ strategy. This is the so-called ``punish, but not too hard'' effect, where too large costs of sanctioning do more damage to those that execute punishment than the imposed fines do damage to the defectors \cite{helbing_njp10}. A direct comparison with the results presented in panel (a) demonstrates clearly that we can observe a similar cooperation among punishing strategies as we have reported before for the low $r$ case, in particular because all the mentioned mildly punishing strategies can survive even at a high $\alpha$ value.

On the other hand, a conceptually different mechanism can be observed in the small $\alpha$ region, which is reminiscent of what one would actually expect from a selection process. More specifically, panel~(a) of Fig.~\ref{r4} shows that at $\alpha \approx 0.2$ only strategy $C_3$ survives and coexists with $D$ while all the other punishing strategies die out. The latter players are those, who could survive individually with defectors but should die out because of the presence of a more effective ($C_3$) strategy. Interestingly, the mentioned selection mechanism can work most efficiently when the leading strategy is less efficient against defectors. Right panel of Fig.~\ref{r4} shows that $C_3$ would be unable to crowd out strategy $D$ at these $\alpha$ values, while a $D$-free state could be obtained at higher $\alpha$ value. In the latter case, when $C_3$ is too powerful, then this strategy beats defectors too fast which allows other punishing strategies to survive: this is similar to what we have observed in the third row of Fig.~\ref{fig3}. But when $C_3$ is less effective at smaller $\alpha$ values then the presence of surviving $D$ players enables $C_3$ players to play out their superior efficiency if comparing to other punishing strategies. Thus, depending on the key parameter values, most prominently the multiplication factor $r$ and the punishment fine $\alpha$, the different punishing strategies can either cooperate with each other or compete against each other in the spatial public goods game.

\section{Conclusion}

We have introduced and studied multiple types of punishing strategies that sanction defectors with different probabilities. The fundamental question that we have addressed is whether there exists a selection mechanism which would result in an unambiguous victor when these strategies compete against defectors. We have shown that the answer to this question depends sensitively on the external conditions, in particular on the value of the multiplication parameter $r$. If the public goods game is demanding due to a low value of $r$, then the pure payoff-driven individual selection provides a helping hand to those punishing strategies that would be unable to survive in an individual competition against defectors. In particular, we have demonstrated that the failure or success of a specific punishing strategy could depend sensitively on the relation of invasion velocities between specific punishing strategies and the defectors. Accordingly, if the loosing punishing strategy can delay the complete victory of defectors sufficiently long, then a more successful punishing strategy has a chance to wipe out defectors first. This is an example of the cooperation between different punishing strategies.

On the other hand, in a less demanding environment, characterized by a higher multiplication factor, a different kind of relation can emerge. While the previously summarized cooperation between punishing strategies is still possible, there also exist parameter regions where competition is the dominant mode, and indeed there is always a single and unambiguous victor among the different classes of punishers. Interestingly, we have shown that this happens when the fittest punishing strategy is not effective enough to beat defectors completely. Instead, by carefully taming the defectors, they help to reveal the advantages of other punishing strategies. As we have shown, the key point here is again the relation between the invasion velocities. Namely, a too intensive invasion will decimate defectors too fast and the advantage of specific punishing classes will remain forever hidden. Therefore, in contrast to intuitive expectation, the social diversity of cooperators in terms of their relations with defectors could be the result of an effective selection mechanism. We hope that this research will contribute relevantly to our understanding of the emergence of diversity among competing strategies, as well to their role in determining the ultimate fate of the population.

\begin{acknowledgments}
This work was supported by the Fundamental Research Funds of the Central Universities of China, the Hungarian National Research Fund (Grant K-101490), the Slovenian Research Agency (Grant P5-0027), and by the Deanship of Scientific Research, King Abdulaziz University (Grant 76-130-35-HiCi).
\end{acknowledgments}

\end{document}